# Fractal Inspired Models of Quark and Gluon Distributions and Longitudinal Structure Function $F_L(x, Q^2)$ at small $x$

Akbari Jahan and D. K. Choudhury

Department of Physics, Gauhati University, Guwahati – 781 014, Assam, India.

## Abstract

In recent years, Fractal Inspired Models of quark and gluon densities at small $x$ have been proposed. In this paper, we investigate longitudinal structure function $F_L(x, Q^2)$ within this approach. We make predictions using the QCD based approximate relation between the longitudinal structure function and the gluon density. As the Altarelli-Martinelli equation for the longitudinal structure function cannot be applied to Model I due to the presence of a singularity in the Bjorken $x$-space we consider Model II only. The qualitative feature of the prediction of Model II is found to be compatible with the QCD expectation.

# 1 Introduction

*Fractal Geometry* [1, 2, 3] defining various fractals occurring in nature has its applications in several areas of physical sciences including Condensed matter physics and solids [4]. Since nineteen eighties the notion of fractals has also found its applicability in high energy physics through the self-similar nature of hadron multi-particle production processes [5, 6, 7]. Specifically during mid-nineties, James D. Bjorken [8] highlighted the fractality of parton cascades leading to the anomalous dimension of the phase space.

However, relevance of these ideas seemed unnoticed in the contemporary physics of Deep Inelastic Scattering (DIS) until 2002 when Lastovicka [9, 10] proposed a relevant formalism and a functional form of the structure function $F_2(x,Q^2)$ at small $x$.

The study of structure functions at low $x$ has become topical in view of the high energy collider like HERA where previously unexplored small $x$ regime is being reached. Gluons are expected to be directly measurable in the small $x$ regime [11, 12]. It is also directly related to the longitudinal structure function $F_L(x,Q^2)$ through the Altarelli-Martinelli Equation [13]. It is therefore vital to have an accurate measurement of $F_L(x,Q^2)$ at HERA since this gives an independent test of the gluon density at low $x$. The experimental determination of $F_L(x,Q^2)$ is difficult, since it usually requires cross-section measurements at different values of the centre of mass energy implying change of beam energies. However, recently, the H1 Collaboration at HERA [14] has extracted experimental results on $F_L$ by measuring the cross-section in a kinematical region where the $F_L$ contribution is substantial. For an extraction of $F_L(x,Q^2)$ it has model-dependent uncertainties which are difficult to quantify fully.

In fixed-target DIS experiments scaling violations have been observed, i.e. the variation at fixed values of Bjorken-$x$ of the structure functions with $Q^2$, the squared four-momentum transfer between lepton and nucleon. These scaling violations of $F_2(x,Q^2)$ are well described by the DGLAP evolution equations [15, 16, 17]. The strong scaling violations observed at low $x$ are attributed to the high gluon density in the proton. In the Quark Parton Model (QPM), $F_2(x,Q^2)$ is the sum of the quark and anti-quark $x$ distributions, weighted by the square of the electric quark charges, i.e.

$$F_2(x,Q^2) = x\sum_i e_i^2 (q_i(x,Q^2) + \overline{q}_i(x,Q^2)) \tag{1}$$

whereas the value of $F_L$ is zero as given by the Callan-Gross relation [18]. However, in Quantum Chromodynamics (QCD), the longitudinal structure function differs from zero, receiving contributions from quarks and gluons. At low $x$ and in the $Q^2$ region of DIS, the gluon contribution greatly exceeds the quark contribution. Therefore $F_L$ is a direct measure of the gluon distribution to a very good approximation.

The Fractal Inspired Model of the nucleon structure function proposed in Ref. [9, 10] has been designed to be valid at small Bjorken-$x$. Later, the model was improved [19, 20] imposing positivity on the "fractal" parameters of the model as well as making it free from singularities in the physical $x$ regime $0 \leq x \leq 1$. A similar model for the gluon distribution was also reported subsequently [21].

The present paper reports an analysis of the longitudinal structure function $F_L(x,Q^2)$ within the Fractal Inspired Models. Specifically we study the status of the relations between $F_L(x,Q^2)$ and gluon density $G(x,Q^2)$ in such models and compare with the recent data [14, 22].

## 2 Formalism

2.1 *Longitudinal Structure Function and Gluon Distribution*

The longitudinal structure function $F_L(x,Q^2)$ comes as a consequence of the violation of Callan-Gross relation [18] in Quark Parton Model and is defined as

$$F_L(x,Q^2) = F_2(x,Q^2) - 2xF_1(x,Q^2) \tag{2}$$

However, in QCD, $F_L \neq 0$ and is given by the Altarelli-Martinelli Equation [13],

$$F_L(x,Q^2) = \frac{\alpha_s(Q^2)}{\pi}\left[\frac{4}{3}\int_x^1 \frac{dy}{y}\left(\frac{x}{y}\right)^2 F_2(y,Q^2) + 2\sum_i e_i^2 \int_x^1 \frac{dy}{y}\left(\frac{x}{y}\right)^2 \left(1-\frac{x}{y}\right) G(y,Q^2)\right] \tag{3}$$

$e_i$ being the electric charge of the $i^{th}$ parton.

As early as 1988, Cooper-Sarkar, Ingelman, Long, Roberts and Saxon [23] obtained the following approximate relation between $F_L(x,Q^2)$, $G(x,Q^2)$ and $F_2(x,Q^2)$ making Taylor series expansion of the integrand of Eq (3) around $x = 0$.

$$F_L(0.417x,Q^2) = \left(\frac{4\alpha_s}{3\pi}\right)\left[\frac{5}{3}\frac{1}{5.8}G(x,Q^2) + \frac{1}{1.97}F_2(0.75x,Q^2)\right] \quad (4)$$

This has been used by several authors [24, 25] to estimate the longitudinal structure function with considerable success.

## 2.2 *Fractal Inspired Models*

**Model I:**

This is the original version of the structure function used in Ref. [9, 10]. The gluon density based on the same set of magnification factors was reported in Ref. [21]. They are:

$$F_2(x,Q^2) = \frac{(\exp D_0)Q_0^2 x^{-D_2+1}}{1 + D_3 + D_1 \log\frac{1}{x}}\left[x^{-D_1 \log\left(1+\frac{Q^2}{Q_0^2}\right)}\left(1 + \frac{Q^2}{Q_0^2}\right)^{D_3+1} - 1\right] \quad (5)$$

$$G(x,Q^2) = \frac{(\exp D_0^g)Q_0^2 x^{-D_2^g+1}}{1 + D_3^g + D_1^g \log\frac{1}{x}}\left[x^{-D_1^g \log\left(1+\frac{Q^2}{Q_0^2}\right)}\left(1 + \frac{Q^2}{Q_0^2}\right)^{D_3^g+1} - 1\right] \quad (6)$$

where the parameters are

$D_0 = 0.339 \pm 0.145$, $D_1 = 0.073 \pm 0.001$, $D_2 = 1.013 \pm 0.01$, $D_3 = -1.287 \pm 0.01$, $Q_0^2 = 0.062 \pm 0.01$ GeV$^2$. (7)

and

$D_0^g = 2.1961$, $D_1^g = 0.073$, $D_2^g = 1.2662$, $D_3^g = -1.287$, $Q_0^{2g} = 0.062$ GeV$^2$. (8)

While the set for quark densities (Eq (7)) is fixed from HERA data [26], set for the gluon densities (Eq (8)) is fixed from MRST results [27].

**Model II:**

With the new set of magnification factors $\left(\frac{1}{x} \text{ and } \frac{Q_0^2}{Q_0^2 + Q^2}\right)$ instead of $\left(\frac{1}{x} \text{ and } \frac{Q_0^2 + Q^2}{Q_0^2}\right)$, the corresponding expressions for the structure function [19] and the gluon distribution [21] are:

$$F_2(x, Q^2) = \frac{(\exp D_0) Q_0^2 x^{-D_2+1}}{1 - D_3 - D_1 \log \frac{1}{x}} \left[ x^{D_1 \log\left(1 + \frac{Q^2}{Q_0^2}\right)} \left(1 + \frac{Q^2}{Q_0^2}\right)^{-D_3+1} - 1 \right] \quad (9)$$

$$G(x, Q^2) = \frac{(\exp D_0^g) Q_0^2 x^{-D_2^g+1}}{1 - D_3^g - D_1^g \log \frac{1}{x}} \left[ x^{D_1^g \log\left(1 + \frac{Q^2}{Q_0^2}\right)} \left(1 + \frac{Q^2}{Q_0^2}\right)^{-D_3^g+1} - 1 \right] \quad (10)$$

where

$$D_0 = 0.6345 \pm 0.0145,\ D_1 = 0.2398 \pm 0.0125,\ D_2 = 1.2581 \pm 0.0157,$$
$$D_3 = 1.4352 \pm 0.0113,\ Q_0^2 = 0.0498 \pm 0.0013 \text{ GeV}^2. \quad (11)$$

and

$$D_0^g = 2.9594,\ D_1^g = 0.2398,\ D_2^g = 1.4484,\ D_3^g = 1.4352,\ Q_0^{2g} = 0.049 \text{ GeV}^2. \quad (12)$$

Note that the parameters $D_3$ and $D_3^g$ of Eq (11) and Eq (12) are positive unlike in Eq (7) and Eq (8).

## 3 Results and Discussions

**Model I:**

Eq (5) has two limitations. First, the parameter $D_3$ is negative contrary to the expectation of positivity of the fractal dimensions. Secondly, due to its negative value, Eq (5) develops a singularity at $x \approx 0.019614$ as it satisfies the condition $1 + D_3 + D_1 \log \frac{1}{x} = 0$, contrary to the expectation of a physically viable form of structure function. Same is the case for Eq (6) as well. It also has negative value for the parameter $D_3^g$ and a singularity for the same value of $x$ as that of Eq (5). Due to the presence of such singularity, calculation of $F_L$ using Altarelli-Martinelli Equation [AM] (Eq (3)) for Model I is not possible. Hence for Model I, calculation of $F_L$ is given only by the Cooper-Sarkar *et al* relation [CS] (Eq (4)).

**Case 1:** Plotting the graph between $F_L(x,Q^2)$ and $x$ [Fig. 1], it is observed that $F_L(x,Q^2)$ increases towards low $x$, which is consistent with the NLO QCD calculation [14] and thus reflects the rise of the gluon distribution in this kinematical region. For each fixed $Q^2$, the experimental H1 data [14, 22] lie beneath the predicted $F_L(x,Q^2)$ vs $x$ graph. As $Q^2$ increases, the H1 data move closer to the theoretical predictions and then almost coincide with it at $Q^2 = 90$ GeV$^2$.

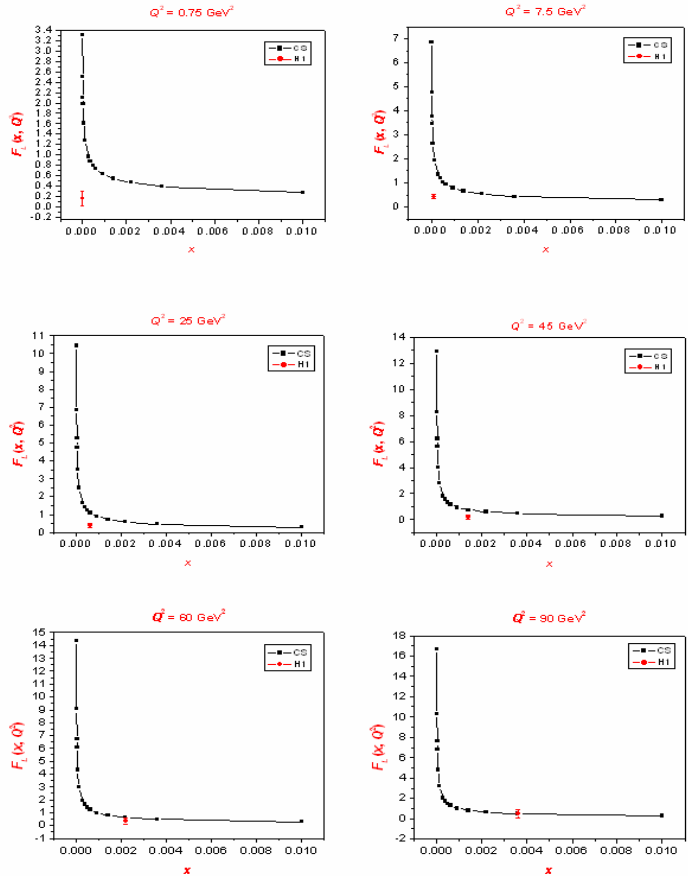

**Figure 1:** $F_L(x,Q^2)$ vs $x$ at $Q^2 = 0.75, 7.5, 25, 45, 60$ and $90$ GeV$^2$ for Model I using Eq (4).

**Case 2:** Keeping *x* fixed, a graph is plotted between $F_L(x,Q^2)$ and $Q^2$ [Fig. 2]. In this case, the predicted $F_L$ rises steadily with $Q^2$. Our theoretical predictions obtained using Eq (4) lie much above the experimental H1 data [14, 22]. As *x* increases ($x \geq$ 0.0022) the predictions come closer to the data.

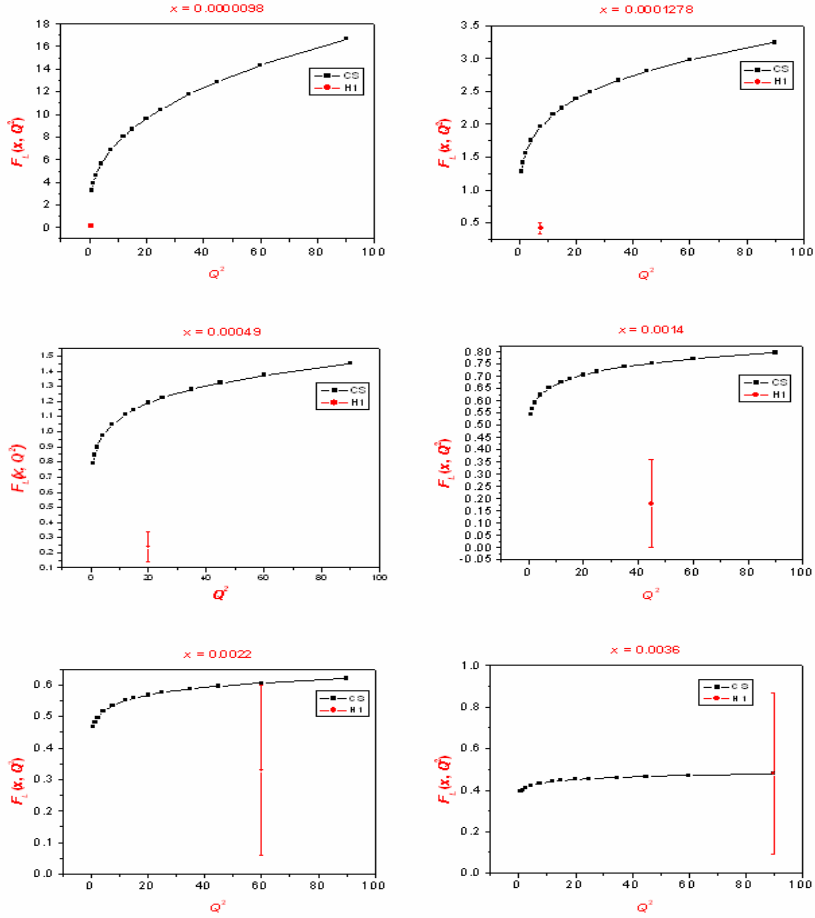

**Figure 2:** $F_L(x,Q^2)$ vs $Q^2$ at *x* = 0.0000098, 0.0001278, 0.00049, 0.0014, 0.0022, 0.0036 for Model I using Eq (4).

**Model II:**

As noted earlier, the problem of singularity is overcome in Model II and thus the Altarelli-Martinelli Equation (Eq (3)) can be used for this Model.

**Case 1:** For each fixed $Q^2$, a graph is plotted between $F_L(x,Q^2)$ and $x$ [Fig. 3]. Here, the H1 data lie beneath the graph (as in the case of Model I) and then at $Q^2 = 25$ GeV$^2$ the H1 data crosses the graph and lies above it for the remaining higher $Q^2$ values.

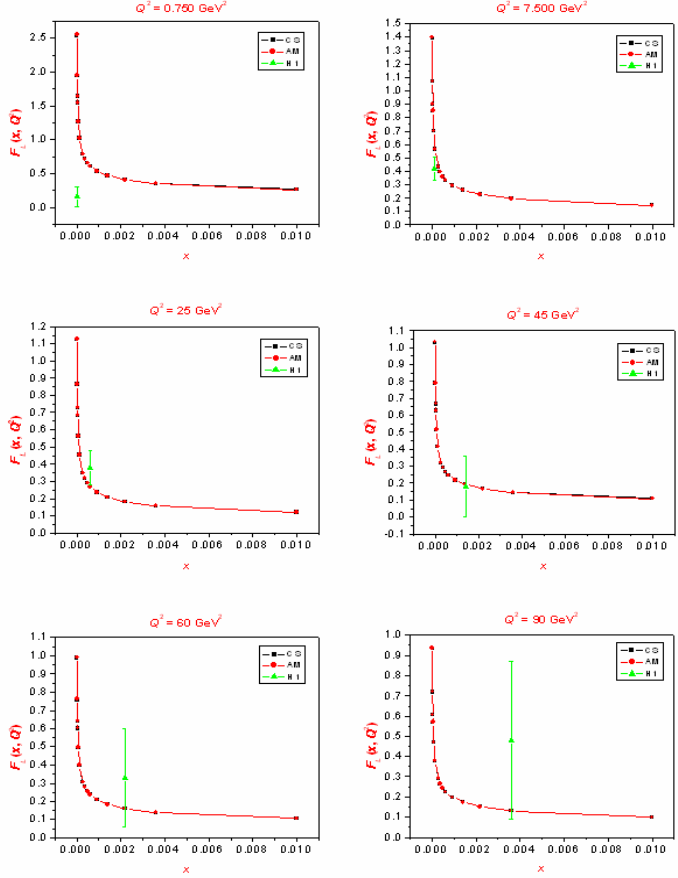

**Figure 3:** $F_L(x,Q^2)$ vs $x$ at $Q^2 = 0.75, 7.5, 25, 45, 60$ and $90$ GeV$^2$ for Model II using Eq (3) and Eq (4).

**Case 2:** Keeping $x$ fixed, a graph is plotted between $F_L(x,Q^2)$ and $Q^2$ [Fig. 4]. In this case, the predicted $F_L$ falls steadily with $Q^2$. Our theoretical predictions obtained using both Eq (3) and Eq (4) lie above the experimental H1 data [14, 22]. As $x$ increases ($x \geq 0.00049$) the predictions come closer to the data. For $x \geq 0.0022$, the data seems to overshoot the theoretical predictions.

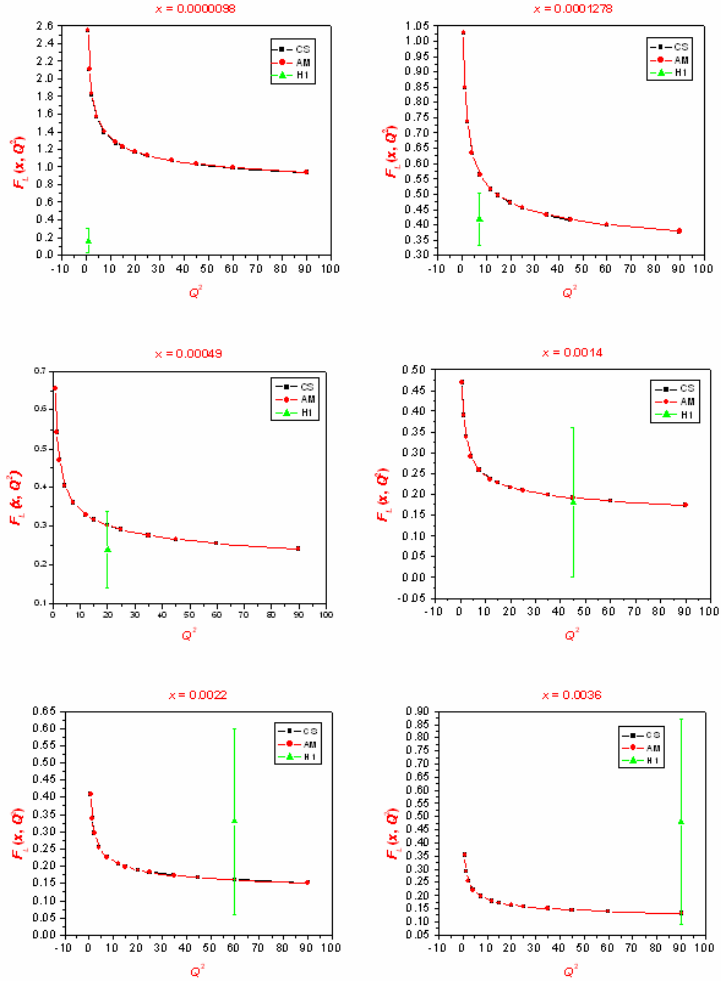

**Figure 4:** $F_L(x,Q^2)$ vs $Q^2$ at $x$ = 0.0000098, 0.0001278, 0.00049, 0.0014, 0.0022, 0.0036 for Model II using Eq (3) and Eq (4).

Thus it can be seen that the experimental H1 data are in moderate agreement with the graphs of Model II. Within the experimental uncertainties the data are consistent with these predictions. This consistency underlines the applicability of the DGLAP evolution framework of perturbative QCD at low Bjorken-*x* for the Fractal Inspired Models under study.

## 4 Conclusion

In this paper, fractality is used as a tool to provide parameterization for the quark and gluon distributions at small $x$. Model I has a singularity at $x \approx 0.019614$. It implies that this parameterization should be used only in the small $x$ regime ($x < 0.019614$) which thus excludes its use in the integration of the Atarelli-Martinelli Equation. We therefore concentrate on Model II only. However the predictions of Model II on $F_L$ using both Cooper-Sarkar *et al* Equation and Altarelli-Martinelli Equation are almost identical implying negligible large $x$ contribution in the Altarelli-Martinelli Equation. While extrapolating the prediction to very large $Q^2$ or very small $x$, we observe very fast growth of gluon densities which might make the longitudinal structure function even exceeding unity (violating the standard relation $0 \leq F_L \leq F_2$). To extend the validity of the present formalism to such kinematical regime, one presumably needs to impose constraint on the gluon distribution due to Froissart bound [28, 29, 30]. Such a possibility is currently under study.

## Acknowledgement

One of the authors (AJ) gratefully acknowledges UGC-RFSMS for financial support.